# Carrier-Envelope-Phase Dependent Dissociation of Hydrogen

**Han Xu (徐晗)[1], J - P Maclean[1], D E Laban[1,2], W C Wallace[1,2], D Kielpinski[1,2], R T Sang[1,2] and I V Litvinyuk[1,\*]**

[1]*Centre for Quantum Dynamics and Australian Attosecond Science Facility, Griffith University, Nathan, Australia 4111*

[2]*ARC Centre of Excellence for Coherent X-Ray Science, Griffith University, Nathan, Australia 4111*

\*E-mail: igor.v.lit@gmail.com

**Abstract.** We studied dependence of dissociative ionization in $H_2$ on carrier-envelope phase (CEP) of few-cycle (6 fs) near-infrared (NIR) laser pulses. For low-energy channels, we present the first experimental observation of CEP dependence for total dissociation yield and the highest degree of asymmetry reported to date (40%). The observed modulations in both asymmetry and total yield could be understood in terms of interference between different n-photon dissociation pathways - n and (n+1) photon channels for asymmetry, n and (n+2) photon channels for yield – as suggested by the general theory of CEP effects (Roudnev and Esry, Phys. Rev. Lett. **99**, 220406 (2007), [1]). The yield modulation is found to be π-periodic in CEP, with its phase strongly dependent on fragment kinetic energy (and reversing its sign within the studied energy range), indicating that the dissociation yield does not simply follow the CEP dependence of maximum electric field, as a naïve intuition might suggest. We also find that a positively chirped pulse can lead to a higher dissociation probability than a transform limited pulse.

Carrier-Envelope-Phase Dependent Dissociation of Hydrogen

The control of electron localization during the laser dissociation of molecules is currently an active area of research due to its potential application in controlling the chemical reaction processes. In homonuclear diatomic molecules such control requires the breaking of inversion symmetry. Several schemes for doing this have been suggested and implemented recently for hydrogen and deuterium molecules: combining NIR pulses with XUV isolated attosecond pulses [2] or attosecond pulse trains [3]; controlling asymmetric waveform of NIR-visible pulses by combining a fundamental frequency with its second harmonic [4]; using few cycle NIR [5] or mid-IR [6] pulses with stable (or measured for each pulse) CEP.

The CEP control with few-cycle pulses was the scheme first demonstrated experimentally by Kling et al. [5] and it continues to be the most extensively studied. In the original study of Kling and co-workers, they demonstrated significant (~20%) CEP-controlled asymmetry of deuteron emission for only the higher energy (KER > 6 eV or > 3eV for individual fragments) dissociation channels, which was attributed to recollision excitation of the dissociative $2p\sigma_u$ state of $D_2^+$. However, no noticeable asymmetry was reported for the dominant low energy channels, i.e. bond softening (BS) and above-threshold dissociation (ATD). More recently, Kremer et al. observed a significant (~15%) asymmetry for these low energy (0.5 eV < KER < 2.5 eV) channels in $H_2$. Neither of these studies reported on CEP dependence of the total dissociation yields including fragments emitted in all directions. We report such measurements of the CEP-dependent total dissociation yield. We also report the highest degree of asymmetry ever measured for dissociation fragments with KER between 1.8 and 3 eV for this control scheme of 40%.

On the theoretical modelling of these processes, extensive efforts have been exerted to understand the underlying physics of the control of electron motion on sub-femtosecond time-scale [8, 9, 10]. For each reported experimental scheme, numerical models solving the time-dependent Schrödinger equations for nuclear motion on two field-coupled electronic potentials (two electronic states of opposite parity are needed for localization), supplemented by an *ad hoc* wavepacket initiation procedure, qualitatively accounted for the observations. More generally, all CEP-dependent effects could be viewed as resulting from the interference between two (or more) quantum pathways corresponding to different numbers of absorbed photons. This general view follows directly from periodic dependence of the time-dependent Schroedinger equation on CEP, which allows Floquet representation of the wavefunction, as was first demonstrated by Roudnev and Esry in [1]. In this interpretation, CEP control (and also the two-color control [4]) is a form of more general coherent control [7].

Two conditions are necessary for achieving coherent control of electron localization: (i) the two interfering pathways must result in electronic states of opposite parity; and (ii) the two interfering pathways must result in the same kinetic energy of fragments. Additionally, to achieve maximum modulation depth, the relative probabilities of the two interfering pathways need to be close (optimally equal) to each other. The condition (ii) may only be satisfied for the regions where kinetic energies of BS, ATD and 3PD overlap. Such an overlap would be hard to achieve with narrowband pulses, as in monochromatic limit the three dissociation processes result in distinctly different kinetic energies. However, the large bandwidth of few-cycle pulses broadens the corresponding spectra and allows such interference to occur (figure 1(a)).

To date both experiments and numerical modelling focused almost exclusively on the asymmetry parameter defined as

$$A = (N_{up} - N_{down})/(N_{up} + N_{down})$$

where $N_{up}$($N_{down}$) is a total number of protons (or deuterons) ejected into each of the two opposite directions along the laser polarization axis. Thus defined asymmetry exhibits periodic dependence on the CEP with $2\pi$ periodicity, which is intuitively obvious from simple symmetry considerations. The general theory of CEP effects predicts a CEP dependence of the asymmetry resulting from the interference of two dissociation pathways which differ by one in the number of absorbed photons [8]. The same theory also predicts that the total dissociation yield ($N_{tot} = N_{up} + N_{down}$) is CEP-dependent, but with $\pi$ periodicity which results from interference of quantum pathways differing by two in number of absorbed photons. In this case numerical modelling predicts only a small (1.4%) modulation in total probability with CEP for $H_2^+$ interacting with 5.9 fs pulses [8]. Detecting such small modulations requires laser pulses which are simultaneously stable in the total energy, pulse duration and CEP. Until now, no experimental results on CEP dependence of total dissociation probability have been reported.

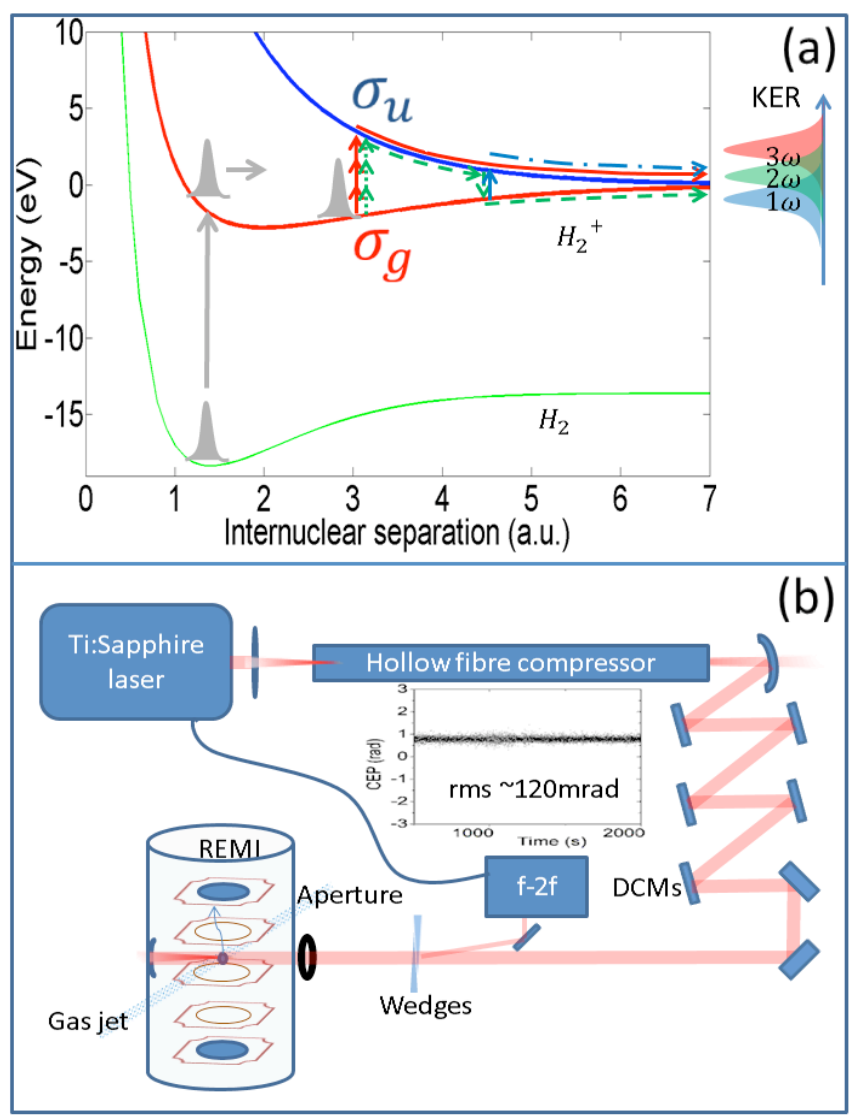

Figure 1. (a) Different pathways for dissociation of $H_2^+$. The arrows represent transitions corresponding to the central frequency of the few-cycle pulse – large bandwidth leads to broad and overlapping fragment energy spectra allowing interference between the pathways. (b) Schematic of the experiment (DCM – dispersion compensation mirrors, REMI – reaction microscope).

Carrier-Envelope-Phase Dependent Dissociation of Hydrogen

We present an experimental study of CEP dependence of dissociative ionization of $H_2$ investigating the low energy (KER < 4 eV) channels. Unlike the high energy fragments observed by Kling et al. [5], here the dissociative $\sigma_u$ state of $H_2^+$ is populated not via rescattering but via radiative excitation pathways corresponding to absorption of odd numbers of photons (n = 1, 3). Bond softening (BS, n = 1) [9] and three-photon dissociation (3PD, n = 3) [10] both populate the $\sigma_u$ state. Above-threshold dissociation (ATD) is a net-two-photon (n = 2) process, where absorption of three photons is followed by emission of one photon resulting in the dissociation on the ground $\sigma_g$ state (figure 1) [11]. We interpret our observations in terms of interference between these three channels.

The schematic of our experiment is shown in figure 1(b). The experimental apparatus includes a 1 KHz CEP-stable few-cycle laser system and a Reaction Microscope (REMI). The few-cycle laser is a commercial femtosecond laser system (*Femtopower Compact Pro, Femtolasers*) with a hollow-core fibre compressor. The laser pulse had duration of 6 fs with central wavelength of 800 nm. A pair of fused silica wedges installed on a translation stage was used to vary the CEP of the laser pulse. In the experiment, we scanned the CEP over a range of $3\pi$ with 14 sampling points. A small portion of the pulse energy, reflected from the front surface of the wedge pair, was diverted into an f-2f interferometer, where the spectrum was measured and used as a feedback signal for a CEP stabilization feedback loop. The measured CEP noise (root mean square) was less than 120 mrad and was sufficiently stable for clear experimental observation of CEP dependent effects. The laser was tightly focused by a silver-coated concave mirror (*f = 75 mm*) installed inside the REMI onto a well-collimated supersonic gas jet of hydrogen molecules. The electric field of the laser was parallel to the time-of-flight axis of the REMI spectrometer and normal to both the laser propagation direction and the molecular beam. The charged dissociation fragments (protons) were detected by a time and position-sensitive detector (Roendtek) and their three dimensional momentum vectors were determined.

From the measured proton momenta we calculated the kinetic energy spectra separately for protons emitted towards (up) and away from (down) the detector for all values of CEP (figures 2(b) and 2(c)). From those spectra we obtained energy- and CEP-dependent asymmetry and total dissociation yield shown in figures 2(d) and 2(e) respectively. Note that we express the kinetic energy in terms of kinetic energy release (KER), the total kinetic energy of the two fragments. The KER is twice the single fragment energy (used in [5]), assuming equal energy sharing between the proton and the undetected hydrogen atom. Clearly, there are two distinct energy regions (0.3-1.3 eV and 1.5-3.5 eV) characterized by different, and nearly opposite, CEP-dependent directionality of proton emissions (figure 2(f)). By properly adjusting the peak laser intensity at the focus (I = $4\times10^{14}$ W/cm$^2$) we achieved a high degree of asymmetry (40%) for fragments with KER between 2 and 3 eV. Changing this optimal intensity in either direction decreased the measured asymmetry. This intensity is significantly higher (by a factor of four) than the one used by Kling and co-workers in [5]. We assign the two energy regions as corresponding to interfering 1- and 2-photon pathways (BS and ATD, 03-1.3 eV) and 2- and 3-photon pathways (ATD and 3PD, 1.5-3.5 eV).

Carrier-Envelope-Phase Dependent Dissociation of Hydrogen

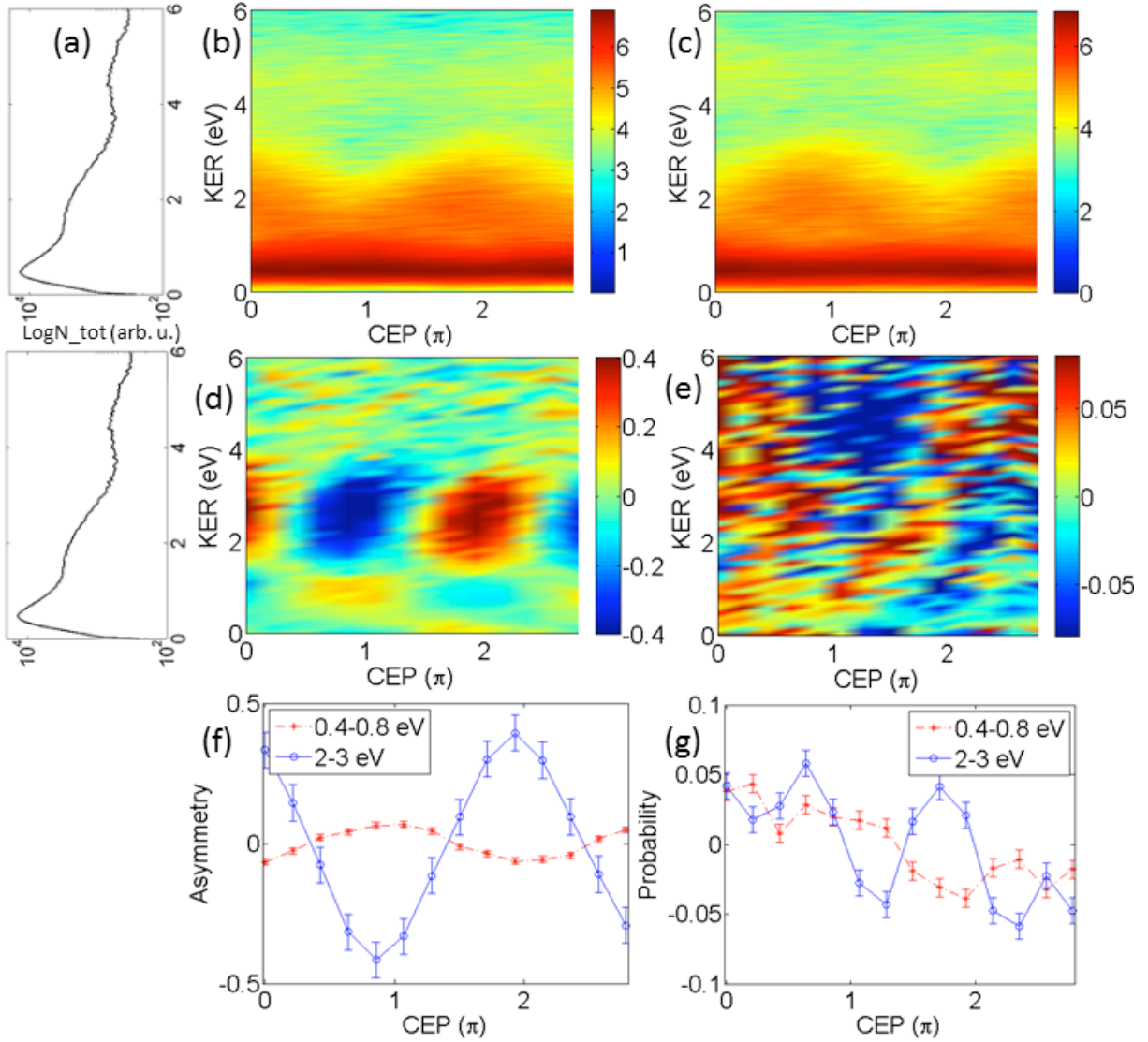

Figure 2. (a) Total KER spectrum (both directions) averaged over all CEP values. CEP-dependent KER spectra for protons ejected towards (b) and away from (c) the detector. (d) Asymmetry as a function of CEP and KER. (e) Total dissociation yield as a function of CEP and KER: colormap indicates the difference between normalized total spectrum at the given CEP and the normalized CEP-averaged spectrum. CEP-dependent asymmetry (f) and relative total yield (g) for two KER regions indicated in the insets.

The highest asymmetry modulation of 40% is achieved for the energy region corresponding to the overlap between the interfering ATD (n = 2) and 3PD (n = 3) channels, reflecting the relative populations of the three dissociation pathways at the experimental intensity of $4 \times 10^{14}$ W/cm$^2$. At the relatively low intensity of $1 \times 10^{14}$ W/cm$^2$ in Kling's experiment [5] it appears that only the BS channel was operational, while the ATD pathway, also required for electron localization, remained unpopulated, thus explaining the absence of asymmetry. It should also be noted, that it will be, in general, harder to achieve CEP control of electron localization in the low energy dissociation channels for heavy (in comparison with light) hydrogen when one is starting with ionization of neutral molecules. The reason for this is the time delay between the initial ionization (typically taking place at the peak of the pulse) and radiative coupling which takes place most efficiently at much larger internuclear separations where the energy difference between the two electronic states is smaller. For very short pulses, by the time the nuclear wavepacket reaches the coupling region the instantaneous pulse intensity is substantially diminished. This is particularly true for heavier isotopes, which move slower and take more time to expand. Experimental comparison of $H_2$ and $D_2$ validates this reasoning, with light hydrogen showing twice the asymmetry modulation depth seen for $D_2$ under the same conditions. The detailed comparison between the two isotopes will be presented in another paper. It is also possible that in this particular case having the shortest pulse may not be the best way of maximizing the CEP effects, as significant intensity at the tail of the pulse is required to populate two different dissociation pathways with comparable probabilities. One may need to optimize both peak intensity and pulse duration to achieve the strongest CEP effects.

We also measured the dependence of total dissociation yield on CEP (figures 2(e) and 2(g)). It is clear from simple symmetry considerations that this dependence must be $\pi$-periodic with CEP. Any given pulse and its mirror image, with a CEP that differs by $\pi$, must show the same total dissociation probability (and opposite asymmetry) in all possible channels. One may naively expect the total yield to just follow the CEP dependence of maximum peak electric field of the pulse, which oscillates between maximum (for cosine pulse) and minimum (for sine pulse) values with a period of $\pi$. This is not what we observe in our experimental results. We do see $\pi$-periodic modulations of total dissociation yield with modulation depth of up to 5% for fragments with KER between 2 and 3 eV. This modulation depth is significantly larger than the theoretical prediction in the work by Hua and Esry for hydrogen molecular ions (1.4%, [8]). At the same time, the relative phase of those modulations is found to be strongly dependent on KER covering more than $\pi$ in range. For instance, when the 2-3 eV fragments are at their maximum yield, the production of 0.4-0.8 eV fragments is minimal (figure 2(g)). We conclude that more complex interference (rather than simple intensity) effects are responsible for CEP dependence of the total yield. More detailed analysis based on the general theory of CEP effects reveals that the yield modulations result from interference between pathways differing by two in number of absorbed photons [8]. In this case the likely interfering pathways are BS (n = 1) and 3PD (n = 3).

Finally, we would like to comment on the monotonic decrease in total ionization yield with increasing CEP, which is seen on top of the periodic modulations in our results (figure 3). That is an artefact produced though the moving the fused silica wedges to alter the CEP since this action simultaneously changes the chirp and duration of the pulse. We placed the transform-limited pulse in the middle of our CEP scanning range and confirmed this by measuring the yield of enhanced (double) ionization fragments [12, 13] which is very sensitive to pulse duration and most strongly suppressed for shortest pulses [14]. It appears that it is not just the duration of the pulse, but also its chirp that affects the dissociation probability. Even a small positive chirp around the point of zero group delay dispersion seems to enhance dissociation probability in comparison to transform-limited and negatively chirped pulses.

Carrier-Envelope-Phase Dependent Dissociation of Hydrogen

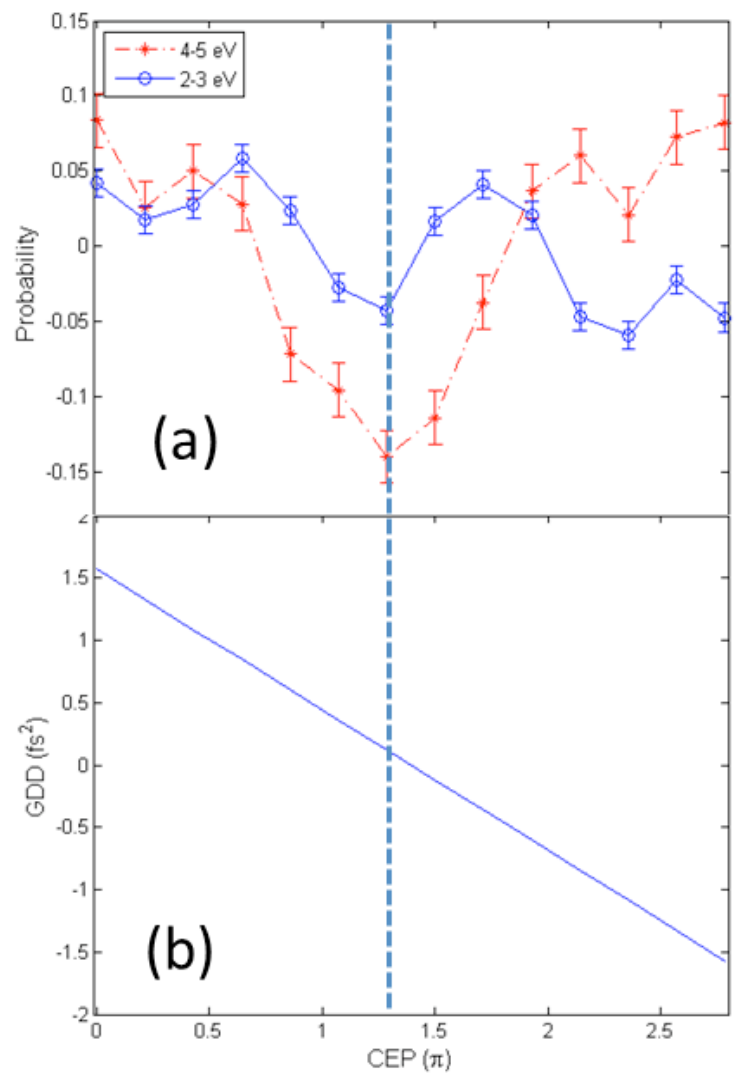

Figure 3. Total yield for fragments with KER in 2-3 eV range and in 4-5 eV range (a) and calculated group delay dispersion (GDD) as functions of nominal CEP. The point of zero GDD is indicated by the vertical dashed line.

In conclusion, we have presented an experimental study of CEP effects in the laser-induced dissociation of hydrogen molecules using 6 fs laser pulses. By optimizing the laser peak intensity we measured the strongest CEP-dependent asymmetry of proton ejection ever reported for this control scheme (40%). This control originates from interference between two distinct dissociation pathways with different number of absorbed photons, above-threshold dissociation (n = 2) and three-photon dissociation (n = 3). We also measured the CEP dependent total dissociation probability, which exhibited π-periodic modulations with up to 5% modulation depth. This dependence originates from the interference between bond softening (n = 1) and three-photon dissociation (n = 3) pathways. We also found that positive chirp favours dissociation for low-GDD few-cycle pulses.

**Acknowledgments.** This work was supported by ARC Discovery Project grant DP110101894 and by the ARC Centre for Coherent X-Ray Science under CE0561787. Han Xu acknowledges support from a Griffith University Postdoctoral Research Fellowship.

Carrier-Envelope-Phase Dependent Dissociation of Hydrogen

**References:**

[1]  Roudnev V and Esry B D, *Phys. Rev. Lett.* **99**, 220406  2007.
[2]  Sansone G *et al. Nature* **465** 763-766 2010.
[3]  Singh K P *et al. Phys. Rev. Lett.* **104**, 023001 2010.
[4]  Ray D *et al. Phys. Rev. Lett.* **103**, 223201 2009.
[5]  Kling M F *et al. Science* **312**, 246–248 2006.
[6]  Znakovskaya I *et al. Phys. Rev. Lett*. **108**, 063002 2012.
[7]  Brumer P and Shapiro M, *Ann. Rev. Phys. Chem.* **43**, 257-282 1992.
[8]  Hua J J and Esry B D, *J. Phys. B: At. Mol. Opt. Phys.* **42**, 085601 2009.
[9]  Bucksbaum P H, Zavriev A, Muller H G and Schumacher D W, *Phys. Rev. Lett.* **64**, 1883 1990.
[10]   Litvinyuk I V *et al. New J. Phys.* **10**, 083011 2008.
[11]   Guistisuzor A, He X, Atabek O and Mies F H, *Phys. Rev. Lett.* **64**, 515 1990.
[12]   Zuo T and  Bandrauk A  D, *Phys. Rev. A* **52**, R2511 1995.
[13]   Seideman T, Ivanov M Yu, and Corkum P B, *Phys. Rev. Lett.* **75**, 2819 1995.
[14]   Legare F *et al. Phys. Rev. Lett.* **91**, 093002 2003.